**Title**: Does car sharing reduce greenhouse gas emissions? Life cycle assessment of the modal shift and lifetime shift rebound effects

**Authors**:
Levon Amatuni [a]

Juudit Ottelin [b]

Bernhard Steubing [a]

José Mogollon [a]

Affiliation address:
(a) Leiden University, PO Box 9500, 2300 RA Leiden, The Netherlands
(b) Aalto University, Department of Built Environment, PL 14100, 00076 Aalto, Finland

E-mail address:
Amatuni L.: l.t.amatuni@cml.leidenuniv.nl
Ottelin J.: juudit.ottelin@aalto.fi
Steubing B.: b.r.p.steubing@cml.leidenuniv.nl
Mogollon J.: j.m.mogollon@cml.leidenuniv.nl

**Highlights**

- Modal shift effect has a strong rebound impact on environmental benefits of CS
- Lifetime shift effect could preserve manufacturing emissions while sharing cars
- LCA approach accounts for such effects if all transport modes are considered
- CS participation reduces annual mobility emissions by 3-18% for an average member

# 1. Introduction

Our planet faces increasing environmental risks imposed by growing rates of greenhouse gas (GHG) emissions toward the atmosphere (Quéré et al., 2018; Urban, 2015). The transportation sector plays a significant role in these greenhouse gas contributions, producing approximately 23% of the global direct $CO_2$ emissions in 2010. In developed economies these contributions are driven by road transportation and become even more significant, reaching circa 30% of their national total (Sims et al., 2014).

There are various approaches to lowering GHG emissions of the mobility sector, and they can be roughly grouped in four categories: *technical*, such as the development of electric vehicles (EV); *legislative*, such as introduction of a carbon or fuel tax; *infrastructural*, such as the development of an extensive urban cycling infrastructure; and *behavioural*, such as promoting vehicle and ride-sharing (Temenos et al., 2017). Car sharing (CS) is a vehicle access scheme, usually delivered by a digital platform, which allows and facilitates communal (shared) rather than private access to a pool of vehicles distributed in the city by a provider such as Car2Go or Zipcar. This should not be confused with on-demand ride-hailing services such as Uber or Lyft (Frenken et al., 2015). In these terms, CS primarily induces behavioural aspects of change.

Recently, CS has gained traction in the urban areas of the developed world, with North America showing a 25% average compound annual member growth rate from 2010 to 2016 (Shaheen et al., 2018a). It has been shown that consumers' image of CS is "greener" than owning a car (Hartl et al., 2018) and that, among others, environmental motives drive the intention to participate in CS (Mattia et al., 2019).

Car-sharing platforms vary significantly in terms of the trip patterns performed by their users, the ownership models, and the stakeholders involved. Nevertheless, members of all types of car-sharing expose two important behavioural effects: 1) change in distances travelled by various modes of transport including their personal vehicles, and 2) change in the vehicle ownership or access patterns (Martin et al., 2010; Martin and Shaheen, 2011; Mitropoulos and Prevedouros, 2014; Namazu and Dowlatabadi, 2018; Nijland and van Meerkerk, 2017; Shaheen et al., 2018b). Such effects could have a strong impact on the GHG emissions related to transportation habits in total. While such sharing practices are frequently advertised and perceived as being inherently more sustainable over private ownership, various rebound effects that could limit these benefits (Frenken, 2017; Schor, 2014) are addressed in this research.

The central aim of this study is to address the effects of CS participation on the transportation habits of an average service user in addition to the corresponding change in GHG emissions. This is achieved via a before-and-after participation comparison using a Life Cycle Analysis (LCA) approach. Contrary to most existing studies on business-to-consumer (B2C) CS participation, this study accounts for GHG emissions related to the modal-shift effect based on real distances travelled by all the modes of transport, that is, the increased use of alternative transport modes caused by decrease in driving. It further includes the non-operational emissions (manufacturing, infrastructure) of all the modes of transport used by the service participants. Finally, it incorporates the automobile lifetime shift effect induced by sharing, i.e. the preservation of the manufacturing rates, during resource sharing, caused by different intensities of usage, the rebound effect which has been rarely addressed in previous studies.

## 2. Materials and methods

2.1 Research design

To estimate such environmental impacts, a before-and-after analysis was conducted comparing total mobility-related emissions for one year before and after car-sharing participation. In this study, these are estimated based on the annual distances travelled by different transport modes and their corresponding cradle-to-grave life-cycle emissions factors. For CS, mode emissions factors were derived from private vehicle factors assuming 3 proposed scenarios for the lifetime mileage (LTM) of shared vehicles.

2.2 Scope of the study

This study considers GHG emissions from the urban mobility sector, expressed in carbon dioxide equivalent ($CO_2$-eq) mass. Three geographical case studies are presented: San Francisco, Calgary, Netherlands. We compare two average CS member's mobility profiles: before and after (during) B2C car-sharing participation in the urban area under consideration. These profiles consist of all the distances travelled by the 5 modes (see Eq.1): private car (*car*), car sharing (*CS*), bus, light rail (tram or metro, *rail*), cycling (*cycle*). Additionally, the 'other', 'walking', and 'carpooling' modes have been considered for some cases. The distances are assumed to be co-dependent, such that an annual reduction in one mode of transport will trigger additional modes of transport. Here, distances travelled by the transport modes have been estimated based on regional transportation statistics and surveys reporting changes in distances travelled by average CS members for each of the three case studies under consideration (San Francisco: Cervero and Tsai, 2007; Calgary: Martin and Shaheen, 2016; Netherlands: Nijland and van Meerkerk, 2017). Even though the Dutch study considers the effects of both B2C and peer-to-peer (P2P) participation and covers the country rather than at the city level, the authors report that the effects are driven mostly by the members of the B2C platforms, which are in practice placed in the urban areas. Our study focuses on the B2C platforms as the impacts of peer-to-peer (P2P) car-sharing platforms on travel behaviour have yet to be statistically quantified.

2.3 Life cycle assessment

LCAs allow for evaluating complete environmental impacts (natural resource depletion or global warming potential, etc.) of a particular product or service considering all the phases of its life cycle (Finnveden et al., 2009). Within the scope of this study, four stages of vehicle life (for each mode independently) were considered (Eq. 2): manufacturing (MANUF), infrastructure (INFR), fuels (FUELS), use (OP) (Figure 1). For each mode's vehicle, the four stages evaluated in this study together contribute at least 90% of the total cradle-to-grave GHG emissions reported by Chester (2008). End-of-life phase was not considered.

In this study, total annual urban mobility-related GHG emissions are estimated using the life-cycle emissions factors for each separate mode and the corresponding annual distances travelled by all the modes, the annual mobility profile which acts as a functional unit of the assessment. See Eq. (1).

$$e_{annual} = e_{PKT_{car}} * VKT_{car} + e_{PKT_{cs}} * VKT_{cs} + e_{PKT_{bus}} * VKT_{bus} + e_{PKT_{rail}} * VKT_{rail} + e_{PKT_{cycl}} * VKT_{cycle} + e_{PKT_{other}} * VKT_{other} \qquad (1)$$

where, $VKT_{mode}$ is the total annual distance travelled by the *mode* (vehicle kilometres travelled) with the corresponding per-passenger kilometre travelled (PKT) life-cycle emission factor given by $e_{PKT_{mode}}$. In the scope of this study, the per-PKT emission factors for each mode are defined as:

$$e_{PKT} = \frac{E_{INFR} + E_{MANUF} + E_{FUELS} + E_{OP}}{LTM * occup} \quad (2)$$

Here, $E_{stage}$ is the total lifetime emissions related to a particular life-cycle *stage* of one vehicle of a specific mode of transport.

Life-cycle emissions factors for a private vehicle, bus, and urban rail (tram or metro) were taken from a study in the United States of America (Chester, 2008) for the four life stages assessed in this study (see Appendix A). Along the study, emissions factors for rail were adjusted according to the local energy grid using corresponding electricity life-cycle emission factors (Appendix B). Life-cycle emissions related to cycling (excluding infrastructure-related emissions) were taken from the European Cyclists' Federation's report (Blondel et al., 2011). Manufacturing-related emissions related to cycling as well as emissions from walking were assumed to be zero. Total life-cycle emissions per-PKT for all the modes, including three possible lifetime mileage scenarios for CS vehicles, are given under Figure 1. Emissions factors for case-specific complementary modes (carpooling, other) were estimated based on the five base modes.

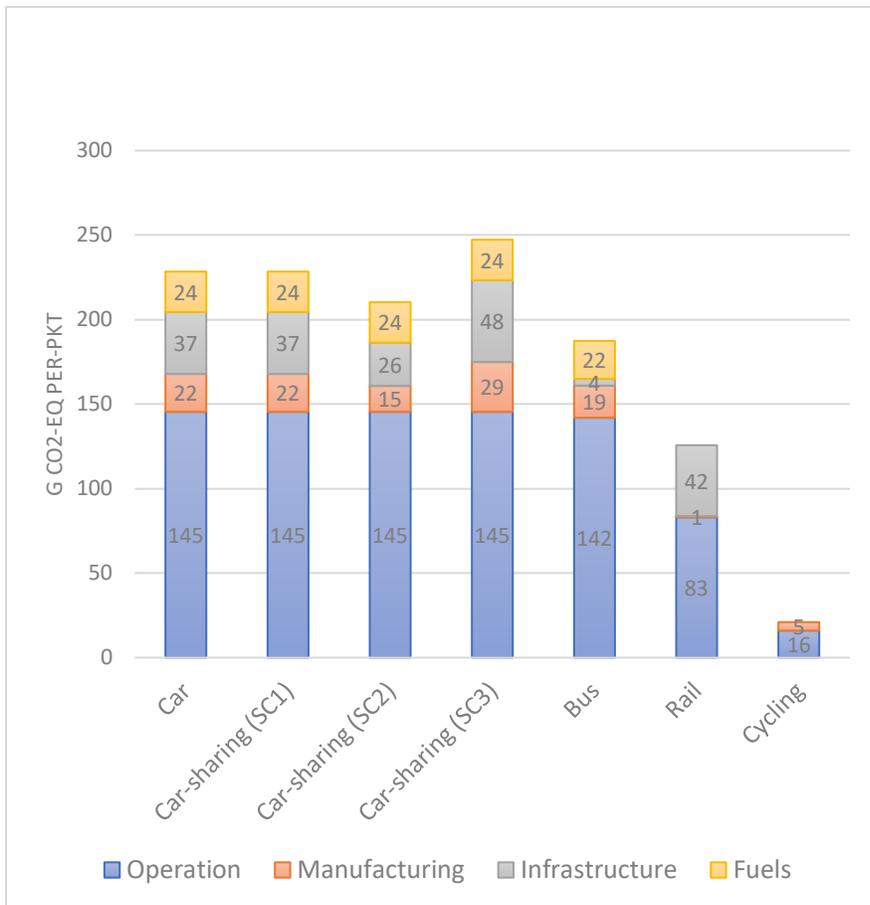

**Fig. 1** Transport modes' emissions factors. Per-passenger kilometre travelled (PKT) emissions (g $CO_2$-eq / km) for various modes of transport based on four selected life stages. Includes three possible car-sharing fleet lifetime mileage scenarios. Electricity emissions factors for Massachusetts, USA has been considered. Data sources: (Chester, 2008; Blondel et al., 2011)

In general, the CS mode's emissions factor was considered to be analogous to that of the private car (including average occupancy), but accounting for the effects of different driving intensity on the CS vehicles' end of life total mileage. Increased access to the same vehicle facilitated by CS services could sharply affect the LTM of that vehicle, which is not usually the case for public transportation. In turn, this could affect the per-PKT emissions of shared cars compared to private cars. Hence, three possible scenarios for the LTM of CS vehicles are considered (see Section 2.4).

After estimating possible lifetime mileage scenarios for CS vehicles, collecting emissions factors for various modes and regions, as well as projecting before-and-after distances for all the modes for three geographical case studies, the total annual emissions reductions induced by CS participation were estimated comparing before and after annual mobility profiles for a given case study using the proposed model with Eq. (1). For specific calculations applied to the various regions see Section 3 and Appendix C in the supplementary materials.

These calculations have been compiled into a graphical user interface for the use of scientific purposes. These are available at the Mobility Emissions Calculator (https://doi.org/10.5281/zenodo.3385074).

2.4 Shared vehicle LTM scenarios

Several studies have suggested faster wear and tear and replacement in shared as compared to privately owned vehicles (Chen and Kockelman, 2016; Meijkamp, 1998). Furthermore, given that shared vehicles are usually sold into the second-hand market and continue their lives as regular personal cars, LTMs for CS are difficult to assess (Meijkamp, 1998). So far, no data has been published on the lifetime mileage of vehicles taking part in car-sharing services. Moreover, based on the methodology from Dun et al. (2015), we have conducted logit regression analysis and observed that both LTM of the vehicle and its lifetime (LT) do not predict its end-of-life, see Appendix D of the supplementary. Due to the lack of data, this study addresses several possible scenarios (Table 1) detailed discussion of which follows this section.

Table 1 Overview of the three proposed scenarios for car-sharing vehicle lifetime mileage given various evidence for the annual mileage

| Scenario number | Scenario | Evidence | Age (years) | Annual Mileage (km) | $LTM_{cs}$ (km) |
|---|---|---|---|---|---|
| 1 | $LTM_{cs} \approx LTM_{car}$ | None | 15 | 16 000 | 240 000 |
| 2 | $LTM_{cs} \gg LTM_{car}$ | Scarce | 12 | 29 000 | 348 000 |
| 3 | $LTM_{cs} \ll LTM_{car}$ | Yes | 15 | 12 200 | 180 000 |

For a meaningful comparison between the scenarios, base lifetime and annual mileage for a private vehicle are set to 15 years and 16000 km accordingly as an average of values from existing sources (see Table G.1 of the supplementary).

Scenario 1: Vehicles which participated in car-sharing at some point in their life do not have a significantly different $LTM_{cs}$ compared to an average private vehicle. This could be the case if these vehicles are not used significantly differently during their car-sharing period, and their average LT stays the same as well. Alternatively, this could be the case if the lifetime decreases (because of the wear & tear) while the mileage increases (because of the higher use intensity) remaining a strong determinant of the vehicle's end-of-life. So far, no literature evidence has been found to support this scenario.

Scenario 2: Average lifetime mileage increases for the car-sharing vehicles due to their more intensified use. This will be the case if the annual mileage increase overweighs the vehicle's shrinking lifetime. Several studies supported this scenario. Meijkamp (1998) suggested that intensified car-sharing use does not allow such age-related causes as corrosion to affect a vehicle's lifetime as fast as its wear & tear, and, as a result, the vehicle reaches its lifetime mileage potential more freely. Further, significantly higher annual mileages (29000 km versus 18000 km for a private vehicle) have been reported once for car-sharing vehicles. However, this data has not been verified by other studies (Mitropoulos and Prevedouros, 2014). A three-year shorter LT has been assumed for this scenario.

Scenario 3: CS vehicles are prone to even lower LTM than their private counterparts. This could be the case if their LT stays the same while annual usage drops because of the CS platform logistics or more driving-conscious CS members being exposed to more explicit participation costs. Moreover, it could be speculated that car-sharing vehicles have a significantly lower lifetime as they are sold to the second-hand market much faster, in around 2 year (Mitropoulos and Prevedouros, 2014), and that this could lower their LTM as well. The first hypothesis was supported after aggregating usage data of the free-floating Car2Go car-sharing service from several North American cities (Car2Go: Press Release, 2018; Martin and Shaheen, 2016) and the CS fleet sizes of these (Car2Go: Pioneer And Market Leader In Free-Floating Carsharing, 2017). This results in relatively lower annual distances of 12200 km (see Appendix E of the supplementary). This scenario follows the results of (Weymar and Finkbeiner, 2016) who argued that smaller automobiles of the lower class (usually the case in the CS fleet), should be assigned lower LTM of around 170000 km.

**3 Case studies**

Before-and after analyses has been conducted separately for three geographical cases: CS members in the Netherlands, City CarShare service in San Francisco, and Car2Go service in Calgary, Canada. The total annual before and after distances travelled are assumed to differ insignificantly as only these CS participants who do not encounter major life events are considered (Nijland and van Meerkerk, 2017), and the total mobility demand is assumed unchanged. A general overview of various dimensions involved in this analysis is presented in Table 2.

Table 2 Overview of the dimensions considered in this study for three case studies. Modes: C = car, CS = car sharing, R = rail, B = bus, Bi = bicycle, W = walking, O = other. Total life-cycle emissions factor has been reported for cycling (Blondel et al., 2011). LCA stages: In = infrastructure, Ma = manufacturing, Op = operation, Fu = fuels (see Section 2.3 and Appendix A). LTM = lifetime mileage (Section 2.4)

| Case-studies | Approach | Transport modes (case specific) | Transport modes (aggregated) | LTM scenarios | Transport LCA stages |
|---|---|---|---|---|---|
| Netherlands | Before-and-after | C, CS, R, B, Bi, CP, O | C, CS, R, B, O | 1 - 3 | In, Ma, Op, Fu |
| San Francisco | | C, CS, R, B, Bi, W, O | | 1 - 3 | |
| Calgary | | C, CS, R, B, Bi, W | | 3 | |

3.1 Netherlands

Nijland and van Meerkerk (2017) showed that the average Dutch car-sharing participant (B2C and P2P platforms averaged) in total drives 1750 km/year less after (during) participating in CS services (7460 instead of 9220) and that 1850 km of that new total driving (private and shared) is done in a shared vehicle. In addition, authors surveyed participants on how their CS-related vehicle kilometres travelled (VKT) would be travelled otherwise, in the absence of the service. The reported car-sharing substitution profile is provided in Table C.1 of the supplementary.

This data along with the total annual distance of 11,000 km reported for an average Dutch citizen (Statistics Netherlands, 2016) allows to estimate the change between annual distances for different transport modes (see Appendix C for details). Given these modal distances and the corresponding emissions factors, annual GHG emissions can be estimated (see Table 3).

Table 3 Estimation of the total 'before' and 'after' annual distances travelled by car-sharing members. Emissions factor of CS is set as a range as it depends on one of the three LTM cases. Total change in annual emissions for each mode based on the projected Dutch car-sharing behavioural scenario. CS emissions change for an unchanged LTM scenario. Italic font for distances signifies those from the original sources rather than estimated here (Nijland and van Meerkerk, 2017)

|  | Before CS (km) | During CS (km) | per-PKT (g $CO_2$-eq) |
|---|---|---|---|
| CS | *0* | *1850* | 210-247 |
| Car | *9220* | *5610* | 228 |
| Train | 1431 | 3069 | 101 |
| Bus | 140 | 299 | 187 |
| Bicycle | 105 | 225 | 20 |
| Carpooling | 35 | 75 | 144 |
| Other | 70 | 150 | 75 |
| Total | 11000 | 11278 |  |

3.2 San Francisco (City CarShare)

Cervero and Tsai (2007) surveyed station-based car-sharing service members in San Francisco and showed that between 2003 and 2005 their daily car-VKT decreased by 38%. Similar to other studies, the authors did not survey the exact change in the distances travelled by various modes; however, they reported around 1609 km of annual car-sharing mileage, constituting 10.1% of the total annual travel distance. In addition to that, the authors' adjacent study reported that rail distances travelled by the CarShare members constituted 33.5% of the total distances travelled, and they surveyed members on the alternative mode choice in the absence of the CS service (Cervero et al., 2007). This data allows to estimate before-and-after annual distances for various modes for this case study (see Appendix C). Table 4 shows these modal distances and the corresponding emissions factors.

Table 4 Before-and-after annual modal distances estimation and emissions comparisons for an average City CarShare member. Emissions factor of CS is set as a range as it depends on one of the three LTM cases. CS emissions change shown for an unchanged LTM scenario only. Italic font for distances signifies those from the original sources rather than estimated here. Sources: (Cervero et al., 2007; Cervero and Tsai, 2007)

|  | Before CS (km) | During CS (km) | per-PKT (g $CO_2$-eq) |
|---|---|---|---|
| CS | *0* | *1609* | 210-247 |
| Car | 9774 | 4451 | 228 |
| Train | 1905 | *5257* | 84 |
| Bus | 1905 | 2331 | 187 |
| Bicycle | 519 | 636 | 20 |
| Walking | 919 | 1125 | 0 |
| Other | 426 | 522 | 125 |
| Total | 15448 | *15931* |  |

3.3 Calgary (Car2Go)

The final case study investigated is based on an existing study of the environmental impacts of a free-floating Car2Go service in five North American cities (Martin and Shaheen, 2016). The authors reported 12,429 VKT driven annually on average by a Car2Go member before car-sharing participation and an average 122 km of annual car-sharing distances for current members in Calgary, Canada. Moreover, an average decrease of 898 km in private vehicle driving was estimated. Such data along with the existing official figures on a complete modal breakdown in Calgary (Behan and Lea, 2014) allow to estimate before-and-after distances for the rest of the modes for this case study (see Appendix C). Table 5 lists the distances projected for all the modes and their corresponding per-PKT emission factors. Contrary to the first two cases where all three LTM scenarios have been considered, here, CS emissions factor corresponds to the reduced LTM Scenario 3 only as it has been specifically estimated for Car2Go fleet (see Appendix E of the supplementary).

Table 5 Before-and-after annual distances and emissions estimated for an average Car2Go member in Calgary. CS emissions change shown for a reduced LTM scenario. Italic font for distances signifies those from the original sources rather than estimated here. Sources: (Martin and Shaheen, 2016)

|  | Before CS (km) | During CS (km) | per-PKT (g $CO_2$-eq) | Emissions (kg $CO_2$-eq) |
|---|---|---|---|---|
| CS | *0* | *122* | 247 | 30 |
| Car | *12429* | *11531* | 228 | -205 |
| Train | 1370 | 1644 | 137 | 38 |
| Bus | 1370 | 1644 | 187 | 51 |
| Bicycle | 571 | 685 | 20 | 2 |
| Walking | 571 | 685 | 0 | 0 |
| Total | 16311 | 16311 |  | -84 |

**4 Results**

Three geographical case studies have been considered: a mix of platforms in the Netherlands, City CarShare in San Francisco, and Car2Go in Calgary. The effect of reduction in total annual urban mobility related GHG emissions caused by participation in the local CS services has been estimated (Figure 2). Pre-CS participation total emissions of the average CS member are compared with their after (during) total emissions to estimate the impact of the platform in the region of application. Complementary transport modes reported within each case study (carpooling, walking, other) have all been aggregated into the 'other' mode, in addition to the five base modes, for a proper comparison.

For the Netherlands, while the reduction in private driving is the strongest single contributor to the change in total emissions (-823 kg $CO_2$-eq), emissions caused by increase in CS driving and other modes moderate the total change significantly. A total annual decrease in 150-219 kg of $CO_2$-eq is estimated (depending on the car-sharing LTM scenario), 186 kg of $CO_2$-eq for the unchanged (middle) LTM scenario. This translates into a 7-10% reduction of the total annual mobility-related emissions because of CS participation. Similar to the Netherlands case, emissions in the San Francisco case study went down because the decrease in driving outweighed the increase of emissions caused by more intensified public transport use. A total decrease of 440 – 500 kg of $CO_2$-eq per member (470 for the middle CS LTM scenario) accounted for a 16-18% decrease relative to the pre-CS participation emissions. For the Calgary case study, we estimate an annual reduction per-member of 84 kg of $CO_2$-eq. According to Martin's study, 84 kg of $CO_2$-eq translates into a 3% reduction of the total transportation-related emissions induced by Car2Go participation of an average member.

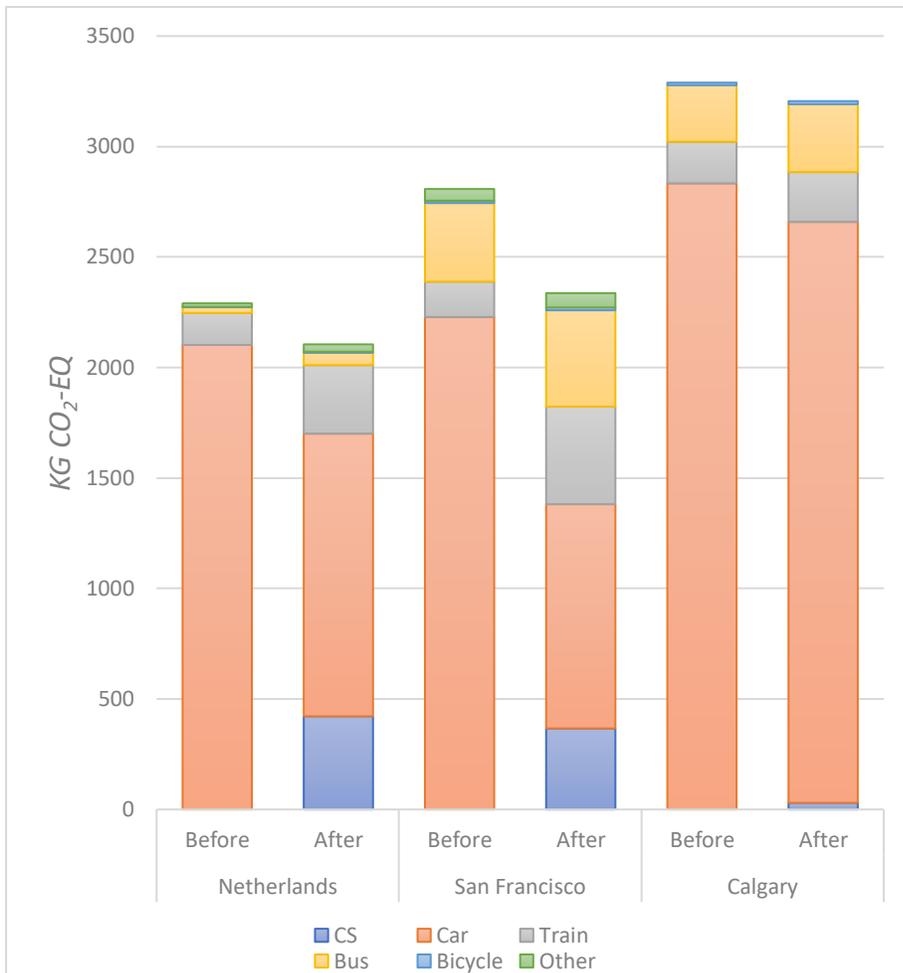

**Fig. 2** Effects of CS on total mobility-related GHG emissions
(before-and-after analysis). Cumulative effect of CS participation on the total mobility-related GHG emissions (kg $CO_2$-eq) for three geographical cases. Before-and-after analysis is based on the projected annual distances for five different transport modes, plus the aggregated 'other' mode for uniformity.

Comparison with the previous results for each region is given in Table 6. Previous CS environmental assessments (Martin and Shaheen, 2016; Nijland and van Meerkerk, 2017) concluded that greater emissions reduction takes place during CS than what our LCA model, Eq.(1), proposes for the same cases.

Table 6 Comparison of the new results with the previous three car-sharing environmental impacts analysis (ranges are based on three LTM scenarios, for Car2Go only decreased 3rd LTM scenario is considered).

| | | Prev. results | New results (LCA) | |
|---|---|---|---|---|
| Case study | Main Data Source | Annual emissions reduction (kg $CO_2$-eq) | Annual emissions reduction (kg $CO_2$-eq) | Annual emissions reduction (rate) |
| Netherlands | Nijland & van Meerkerk, 2017 | -236 to -392 | -150 to -219 | 7 – 10 % |
| San Francisco | Cervero & Tsai, 2007 | N/A | -440 to -500 | 16 – 18 % |
| Calgary | Martin & Shaheen, 2016 | -120 | -84 | 3% |

As a result, based on the regional mobility statistics and existing behavioral surveys, it is estimated that, depending on the regional implication of the CS service, cumulative decrease rate of life-cycle mobility-related GHG emissions varies between 3-18%.

## 5 Discussion

5.1 Comparison of the results

CS has been previously evaluated to incur GHG emissions savings induced by participation in such services (Martin and Shaheen, 2016; Nijland and van Meerkerk, 2017). However, these studies did not take into account the rebound effects of using other forms of mobility while decreasing driving and of the possibly shifting lifetime (or lifetime mileage) of the shared vehicles. In all 3 regional studies, comparing the annual distances travelled by urban modes of transport (before and during CS participation) and applying life-cycle rather than use-related emissions factors, we have found that these rebound effects along with the LCA perspective significantly decrease the GHG emission savings presented in previous work (Table 6).

Furthermore, comparing our results with two existing LCA-based research on environmental impacts of CS, even more significant difference between the results is observed (Table 7). Chen and Kockelman (2016) concluded that an average 50% GHG emissions reduction may occur under CS participation, whereas Ding et al. (2019) predicted up to 37% global warming potential (GWP) reduction due to short-term replacement of private cars.

Table 7 Comparison of the new results with the previous LCA studies on environmental impacts of car sharing.

| LCA study | Reduction in GHG reported | Comments |
|---|---|---|
| this | 3 - 18 % | |

| Ding et al., 2019 | 1 - 37 % | Compared with electric and hybrid vehicles, and carpooling |
| Chen and Kockelman, 2016 | 33 - 67 % | Meta-analysis – combines and averages existing results |

While our findings of 3-18% reduction reported here are significantly lower, they can be attributed to the simultaneous treatment of three important phenomena induced by car-sharing behaviour at least one of which has not be addressed in the listed studies: the modal shift effect, non-operational emissions related to mobility, and the shared vehicle lifetime effect.

5.2 Modal shift effect

Most of the existing studies focused exclusively on the change in driving in their assessment of the environmental impacts. Cervero et al. (2007) surveyed station-based car-sharing service members in San Francisco and showed that between 2003 and 2005 their daily car-VKT decreased by 38%. Similarly, a 2010 report concluded with an impressive decrease in GHG emissions based exclusively on the average 31% decrease in driving and impressive replacement of each 15 personal cars by only one car-sharing vehicle (Frost & Sullivan, 2010). Other studies surveyed Car2Go users by asking them if they changed their usage of other modes of transport after they started using the car-sharing platform (Martin and Shaheen, 2016, 2011). While behavioural changes in VKT driven after car-sharing participation were measured in cardinal values (exact distances), the answers proposed for other modes were based on the ordinal scale (no change, increased, decreased). The authors have found that there was no significant reported change in public transport use on average and accounted for no significant effects of such. On the contrary, real distances travelled by other modes of transport before they had been replaced by car-sharing kilometres have been assessed by Nijland and van Meerkerk (2017), see Table C.1 of the supplementary; however, their study does not take into account that the reported annual decrease in 1,750 kilometres driven by CS members could be replaced by distances travelled by other modes. The LCA study by Chen and Kockelman (2016) is a meta-analysis and is based on most of the aforementioned assessments. Thus, this study, contrary to the previous assessments, assumes and accounts for a simultaneous increase in distances travelled by the alternative modes, which could explain more conservative results observed here. To illustrate the cumulative impact introduced by the modal shift effect, we compare our results (middle LTM scenario) for the three case studies against an assumption where non-driving transportations modes are not considered (Table 8). It is evident that the distance-based modal shift effect considered in this study significantly influences the estimation of the total annual impacts of CS participation and should be considered in the future studies.

Table 8 Comparison of the resulting annual GHG emissions reduction with such if the modal shift effect would not be taken into account (for three case studies)

| Case study | Annual emissions reduction (kg $CO_2$-eq) | Annual emissions reduction w/o modal shift (kg $CO_2$-eq) |
|---|---|---|
| Netherlands | -186 | -401 |
| San Francisco | -471 | -847 |
| Calgary | -84 | -175 |

5.3 Non-operational emissions and the LTM

Non-operational stages (vehicle manufacturing, transportation infrastructure, etc.) of a vehicle's life contribute significantly to the total GHG emissions related to transportation (Chester and Horvath, 2009). However, most car-sharing studies do not take the GHG emissions of these stages for all the modes of transport accessed by the respondents and focus mainly on the lower car manufacturing from fewer owned (Ding et al., 2019; Jung and Koo, 2018; Nijland and van Meerkerk, 2017). Moreover, Nijland and van Meerkerk (2017) attributed all the shed vehicles (sold before the car-sharing membership) to the decrease in manufacturing emissions. On the contrary, our study takes into account that the majority of shed vehicles (sold during CS participation) by new CS members were actually close to their end of life (EoL) already (Martin et al., 2010). Hence, comprehensive treatment of the non-operational emissions for all the modes on a per-use (km) basis proposed here could explain lower emissions reduction rate obtained as well. The model presented in this study automatically assumes that along with those of the private cars, participants are similarly responsible for some portion of the non-operational GHG emissions behind other (public) modes of transportation that they use and whose infrastructure they collectively 'own' and stimulate. Nevertheless, the additional car-sharing related infrastructure emissions were not considered (additional car-sharing stations, web-platform, etc.). The exact non-operational portion of emissions is driven by vehicle's LTM.

Exclusively positive environmental impacts induced by a more frequent shared vehicle replacement have been suggested (Chen and Kockelman, 2016; Meijkamp, 1998) due to better fuel efficiency of the newer cars. One recent study, however, considered the lifetime effect for CS manufacturing emissions admitting significantly lower emissions reduction in Beijing if long-term perspective is taken (Ding et al., 2019). That study, however, assumed a vehicle lifetime and an LTM of 30 years and 600 000 km, respectively. In contrast, Mont (2004) reported that car-sharing vehicles are usually sold to private owners in 2-3 years after being in shared use. Mitropoulos and Prevedouros (2014) suggested that this is closer to 1-2 years and reported 29000 km yearly mileage for the shared vehicles versus 18200 km for the average private vehicle in the US assuming the same 10.6-year average lifetime. Other than that, Oguchi and Fuse (2015) in their study showed that a vehicle's LT varies significantly from country to country and, hence, should be considered for the LCA of such for different geographical regions.

It is not possible to compare the influence of the last two effects (non-operational emissions and lifetime shift) on the total emissions reduction estimations in case of their absence within the scope of this study as they are the inherent elements of the LCA model itself. Nevertheless, even though the total reduction is not significantly sensitive to the proposed LTM scenarios (see Table 6), introducing this parameter into the assessment by itself restricts the positive manufacturing impacts claimed by the previous studies (Chen and Kockelman, 2016; Nijland and van Meerkerk, 2017). Here, we assume that the per-km use of the automobile (both private and shared) induces corresponding portion of such emissions rather than the mere ownership of the vehicle. Considering the lifetime rebound effect, in case if the total annual driving demand stays similar, merely switching from private ownership to a CS scheme does not introduce any non-operational (manufacturing, infrastructure) GHG reductions, as the shared vehicles' lifetimes will be shrinking accordingly, preserving production rates in the long-term.

Comparison of the driving-related non-operational emissions reduction for the Dutch case study (26 to 52 kg of $CO_2$-eq) in our study with the same kind of reduction (125-281 kg of $CO_2$-eq) claimed by the original study (Nijland and van Meerkerk, 2017) exemplifies the impact of the lifetime shift effect on such estimations. This does not consider several other manufacturing-related misalignments in their study (Appendix F of the supplementary).

## 5.4 Sensitivity Analysis

This study evaluated the effects of individual factors such as public transport occupancy levels and local energy production profiles on the GHG emission reductions.

The total annual emissions of an average Car2Go member (section 3.3) are highly sensitive to the occupancy levels of public transport they use to substitute a decrease in driving. For instance the average diesel bus from the US with a 10.5 passenger occupancy was considered while its occupancy ranges from on average 5 to 40 passengers during the day (Chester and Horvath, 2009). Corresponding per-PKT emissions range between 394 and 49 g of $CO_2$-eq. The resulting total mobility-related emissions change would vary from a decrease of 27 kg of $CO_2$-eq for a low-occupancy bus to a 121 kg of $CO_2$-eq emissions decrease for a high-occupancy bus (if the service and members' mobility habits would stay constant otherwise), see Figure 3.

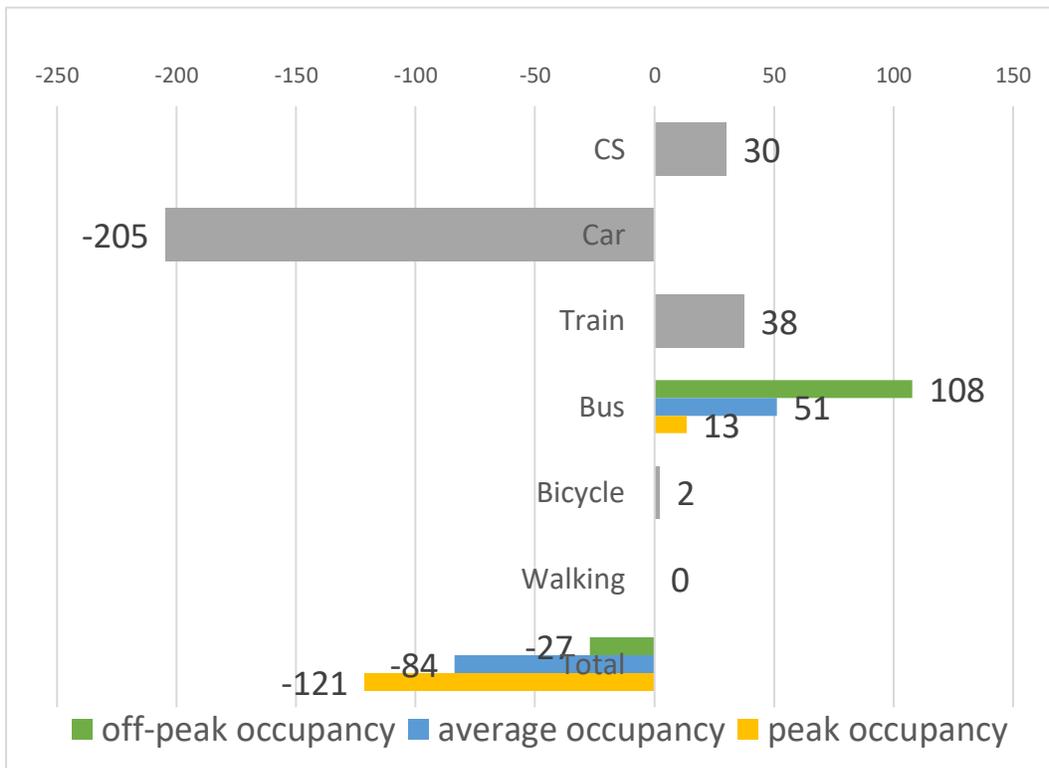

**Fig. 3** Sensitivity to bus occupancy. Total GHG emissions reduction (kg CO2-eq) of the Car2Go car sharing users in Calgary.

In some cases, car-sharing activities could result in even higher total annual mobility-related emissions. For instance, while life-cycle electricity-related emissions factor for California was used, the total annual emissions of an average CarShare member are highly sensitive to the local electricity grid, which powers the trains and the required infrastructure (section 3.2). The resulting total emissions change would vary from a decrease of 663 kg of $CO_2$-eq for a hydroelectric powered Vermont to a 250 kg of $CO_2$-eq emissions increase in an oil and gas sourced electricity grid in Washington D.C, assuming the service and members' mobility habits would stay constant otherwise (see Appendix B of the supplementary materials). Figure 4 depicts such sensitivity analysis.

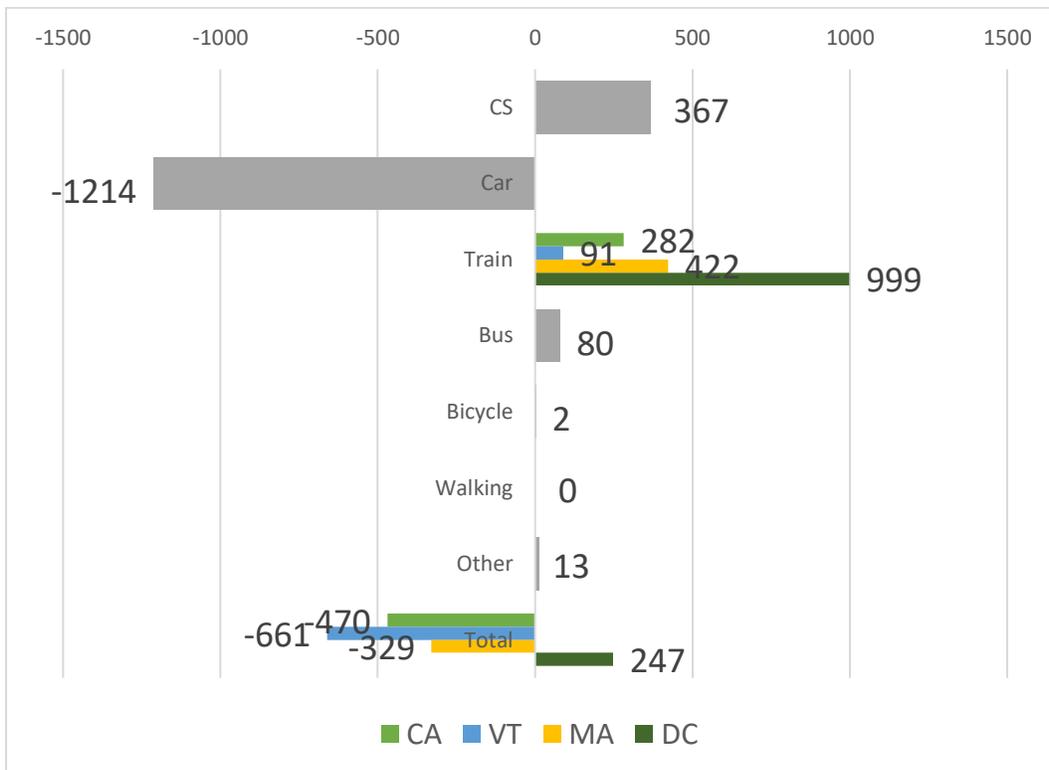

**Fig. 4** Sensitivity to electricity grid. Total mobility related GHG emissions change (kg $CO_2$-eq) of the City CarShare users in San Francisco. Sensitivity to electricity grid in four different states in the USA (California, Vermont, Massachusetts, D.C.)

## 6 Limitations

Even though the socio-transportation system under consideration is highly dynamic, the LCA approach used in the study is inherently *attributional* (Jones et al., 2017). This implies that it is a static snapshot of the system at a particular moment in time and that it operates with the average values for the phenomena under consideration; that is, the total values are divided by the total number of functional units. For instance, the total manufacturing emissions attributed to an average single vehicle or the average emissions related to one additional km travelled by a mode of transport. A different, *consequential* LCA approach would rather consider marginal costs (diminishing returns) caused by an additional km travelled by a particular mode. However, it has been argued that such an approach is not usually feasible for complex transportation systems (Chester, 2008). Still, as this study aims to compare the environmental implications of different mobility habits within an otherwise constant transportation system (rather than system-wide marginal implications caused by changing behaviour), a simplified attributional approach is reasonable. Hence, the impacts of the total amount of the vehicles and the intensity of their usage on the existing infrastructure, fleet production, and maintenance were not considered. Differences between private and shared automobile fuel efficiency levels were not taken into account as well.

Still, it is important to notice that the model under consideration automatically accounts for the *direct rebound effects* (consequences) associated with CS participation. This is the case since it tracks changes in distances travelled by all the modes, and any unintended increase or decrease in the intensity of mobility incurred by an introduction to the eco-efficient innovation is explicitly included in the calculation (Frenken et al., 2017). Hence, our assessment holds characteristics intrinsic to consequential LCA as it evaluates impacts of a changing behaviour (compares mobility profiles) as well. Nevertheless, *the indirect rebound effects* associated with possibly increased or decreased consumption in other consumption categories caused by changing total costs of mobility were not considered in this study (Hertwich, 2005). For example, a study by Ottelin et al. (2017) revealed that reduced car-ownership can lead to significant rebound effects, particularly because of increased air travel. It could be speculated that if the annual costs of car sharing and the substituting modes of transport are lower than the average costs of car ownership and use, the indirect rebound effect would be positive (undesirable), meaning that there are actually additional emissions due to increased consumption in other consumption categories.

Distances travelled annually by an 'average' CS member are considered in this study. Nevertheless, individual annual mobility habits of the CS members could vary significantly. Thus, the mobility-related GHG emissions' 3-18% decrease reported in this study represents the cumulative impacts of such services rather than individual impacts of their members. Hence, for more accurate estimation of such rather personal impacts, annual distances travelled by various modes by the interested individual have to be used in the Mobility Emissions Calculator (https://doi.org/10.5281/zenodo.3385074).

Additionally, citizen-wide surveys and proxies have been used in this study to estimate the distances travelled by all the modes for the before-and-after analysis, whereas, ideally, a survey on the distances travelled by the B2C car-sharing members specifically could provide a better assessment. Nevertheless, the central assumption in the proposed model (that was not followed in the previous studies) was that CS members do not significantly change their grand total annual transportation distance because of CS participation, and this consideration by itself brings in the greatest correction to the previously reported results. Thus, the total emissions reduction is not envisioned to alter significantly if the surveyed modal mix itself differs.

Another point to acknowledge is that, except country-specific electricity emissions factors, USA-based LCA transportation emissions factors are applied for the Canadian and Dutch case studies as well, even though the local transportation systems are different. This could be justified with several reasons. Firstly, the source study for those factors (Chester and Horvath, 2009) is still one of the most comprehensive assessments in the field of LCA for transportation as it includes not only the vehicle and fuel cycles but the infrastructure-related emissions for various transport modes as well. Secondly, applying emissions factors from several national studies with different methodologies could introduce difficult-to-measure distortion to the uniformity of the results. Finally, it seems logical to assume that the major underlining transportation manufacturing and infrastructure-related technologies in those countries are still very similar and that the real differences would not strongly affect the results.

Finally, it has been assumed in this study that occupancy of the shared and private automobiles are the same as it has been reported for the free-floating CS services (Ding et al., 2019). However, possible differences between various types of CS platforms and the corresponding average occupancy could affect the total per-PKT emissions factors significantly and may require further investigation.

## 7 Conclusions

A comprehensive LCA-based model Eq. (1 – 2) has been proposed to estimate the change in the total annual mobility-related GHG emissions caused by average B2C car-sharing participation in three regions (Netherlands, San Francisco, Calgary). For that, life-cycle emission factors for various modes of transport, including region-specific electricity emissions factors, have been used for private automobiles, bus, urban rail (tram, light rail), shared vehicles, and bicycles. To account for a lifetime shift effect induced by sharing, three different emission factors have been considered for a shared automobile given three possible lifetime mileage estimations of such. Moreover, before-and-after participation distances for all the modes under consideration were projected based on the existing data to properly account for the modal shift effect and to calculate total GHG emissions related to transportation habits of an average CS member.

Three case studies considered in this study resulted with a 3-18% reduction of mobility-related life-cycle GHG emissions caused by B2C car-sharing participation by the average member. For all the case studies, the behavioural change in driving had the most significant magnitude of change on the total emissions. The introduction of the modal shift and lifetime shift travel-related rebound effects limit the benefits of decreased driving on GHG emissions. Moreover, the environmental impacts were shown to be highly sensitive to the other characteristics of the transportation system surrounding a particular car-sharing service area: average occupancy of the modes and the electricity grid in the area of application. Interestingly, in rare cases, the total annual mobility-related emissions could even increase if the driving substitution modes are even more carbon-intensive than driving (section 5.4).

This implies that merely using CS vehicles over the private ones does not introduce significantly lower total emissions if the total PKT demand for driving remains constant. On the other hand, it could be argued that given constant PKT distances demand, a ride-sharing or carpooling (versus CS) behaviour (higher automobile occupancy levels) would introduce much more significant reduction in per-PKT emissions factors and the total mobility-related emissions.

Hence, main policy implications should be directed towards reduced automobile use rather than ownership redistribution (sharing) of the vehicles per se. This could be achieved by stimulating use of public modes of transport (including ride-sharing) via corresponding legislative (taxation) and infrastructural measures. Moreover, future research on the environmental impacts related to CS should improve upon the limitations present in this study conducting complete multi-modal distance-based surveys, applying actual local life-cycle environmental impacts factors, as well as incorporating possible technics to account for indirect rebound effects (combined Life Cycle Cost and LCA assessments).

## Acknowledgements

Steve Easterbrook (University of Toronto, supervisor), School of Graduate Studies of the University of Toronto for research assistantship and funding, Tales Yamamoto (Leiden University)

Declarations of interest: none

**Supplementary materials**

Appendix A: Life cycle inventory

Emissions factors for car (including CS), bus, and rail used in this study are obtained using those from the study by Chester (2008). Out all of the life cycle stages considered in the original study, only the major ones, which would contribute to at least 90% of the total emissions factor have been selected for various modes (see Tables A.1 – A.3).

Table A.1 LCI (life-cycle inventory) for automobile used in our study given emissions factors from the original study: Chester (2008); VKT = vehicle km travelled

| LCAI | Emissions factor (kg $CO_2e$ /VKT) | Life-cycle component |
|---|---|---|
| Infrastructure construction | 14000 / $LTM_{car}$ | Roadway Construction |
| Vehicle manufacturing | 8500 / $LTM_{car}$ | Manufacturing |
| Fuel production | 0.038 | Refining & Distribution |
| Vehicle operation | 0.230 | Operation (Running) |

Table A.2 LCI (life-cycle inventory) for a bus used in our study given emissions factors from the original study: Chester (2008); VKT = vehicle km travelled

| LCAI | Emissions factor (kg $CO_2e$ /VKT) | Life-cycle component |
|---|---|---|
| Infrastructure construction | 0.042 | Roadway Construction |
| Vehicle manufacturing | 0.199 | Manufacturing |
| Fuel production | 0.236 | Refining & Distribution |
| Vehicle operation | 1.491 | Operation (Running) |

Table A.3 LCI (life-cycle inventory) for urban rail used in our study given emissions factors of the Green Line Vehicle from the original study: Chester (2008). MJ = megajoule, VKT = vehicle km travelled

| LCAI | Emissions factor per VKT | Life-cycle component |
|---|---|---|
| Infrastructure construction | 1.081 kg $CO_2$e | Station Construction; Track/Power Construction |
| Infrastructure operation | 8.9 MJ | Station Lighting; Station Escalators; Station Train Control; Station Parking Lighting; Station Miscellaneous |
| Vehicle manufacturing | 0.038 kg $CO_2$e | Manufacture |
| Vehicle operation | 30 MJ | Operation (Active); Operation (Idling); Operation (HVAC) |

Later, for our study, these have been adjusted given occupancy rates and automobiles' LTMs in our study to obtain total per-PKT emissions factors instead.

Emissions factor for cycling has been taken from the study by Blondel et al. (2011), and that covers all of the four stages considered for the previous modes except the infrastructure-related emissions which were considered as zero in this study.

Appendix B: Energy life cycle emissions factors

The emissions behind infrastructure and vehicle operations are dependent on a particular energy source profile in the region of operation. For the Canadian provinces of Alberta (AL) this is reported by the National Energy Board of Canada (2016), for the Netherlands (NL) by the Energie Beheer Nederland (2018). See Table B.1.

Table B.1 Energy sources profiles. Sources: NEB Canada, EBN Nederland

| Province | Uranium | Coal | Hydro | Natural Gas | Wind | Solar | Biomass | Petroleum |
|---|---|---|---|---|---|---|---|---|
| AL | - | 47% | 3% | 40% | 7% | - | 3% | - |
| NL | 1% | 14% | - | 40% | 1% | 1% | 4% | 39% |

Given the energy sources profile for a particular region, the following full life-cycle emissions factors for energy production and distribution by different energy sources are used to calculate the average emissions factor for each region (Schlomer et al., 2014). Petroleum's emissions were not found and were taken as for the biomass (Table B.2).

Table B.2 Full-fuel-cycle electricity emissions factors. Source: Schlomer et al., 2014

|  | Nuclear | Coal | Hydro | Natural Gas | Wind | Solar | Biomass | Petroleum |
|---|---|---|---|---|---|---|---|---|
| gCO2eq/kWh | 12 | 820 | 24 | 490 | 11 | 44 | 230 | 230 |

As a result of multiplying these two data sets, the average emissions factor for Alberta weighted by its energy profile is 590 gCO2 eq. per a kWh of energy used. For the Netherlands this comes to 410 gCO2 eq per kWh of energy.

For the US states, energy life-cycle emissions are taken from the previous report (Leslie, 2014): 538 gCO2 eq for a kWh of energy produced in Massachusetts, 327 gCO2 eq in California, 39 gCO2 eq in Vermont, 1397 gCO2 eq in Washington D.C., and 157 gCO2 eq in Washington state.

Appendix C: Region-specific calculations

Netherlands

First, the total annual transportation distance of 11,000 km was assumed as it has been reported for an average Dutch citizen (Statistics Netherlands, 2016). Next, the modal preference profile reported by CS participants (Table C.1) has been used to estimate not only distances travelled by an average member before participation (instead of CS driving) but to estimate distances travelled by alternative modes as total driving decreases after participation as well. This was done, first, redistributing the non-driving 'before' distances between other modes proportionally to the preference profile and, secondly, redistributing the annual distance gap between the 'during' alternative modes ('before' and 'during' distances in Table 3). As a result, total annual distances are not equal because of the effect of the 'would not be travelled' option in the surveyed preference profile. Next, differences in the total distances travelled by each mode were calculated and multiplied by the corresponding per-PKT emission factors to estimate the total annual reduction in emissions caused by CS participation (Table 3).

Table C.1 Car-sharing substitution in the absence of the CS service in the Netherlands. Here, we assumed 'bus' distances as 4% and 'rail' distances as 41% from the original study. Source: (Nijland & van Meerkerk, 2017)

| Mode of transport | Kilometres (in %) |
|---|---|
| Car | 34 |

| Train | 41 |
|---|---|
| Bus, tram, rapid transit | 4 |
| Bicycle | 3 |
| Car passenger | 1 |
| Other | 2 |
| Not travelled | 15 |

Here, rail per-PKT emissions were recalculated using the appropriate electricity emissions factor for the Netherlands. Car passenger (ride-sharing or carpooling) per-PKT emissions were taken as regular car emissions adjusted for occupancy of 2.5 rather than 1.58. Per-PKT emissions for 'other' were averaged across other modes and halved to account for walking (zero emissions). Car-sharing per-PKT emissions were set as a range based on the three CS lifetime mileage scenarios.

San Francisco

Data (several distances and the percentage changes) reported by the original study allows to estimate the total annual 'during' and, consequentially, 'during' rail distances. The rest of the 'during' modes are estimated redistributing the rest of the total distance between other modes using reported preference profile (Table C.2 of the supplementary). Afterward, the 'before' car-VKT could be estimated using the reported decrease rate. Finally, the before-and-after annual distance gap would be redistributed to the 'before' distances of the rest of the modes according to the same preference profile. These were used to estimate before-and-after distances for all the modes using the surveyed modal preference profile to redistribute annual distance gap between unknown distances by alternative modes. Resulting changes in annual distances are in turn multiplied by the per-PKT emission factors of the corresponding modes to obtain total emissions change for each transport mode. Per-PKT emissions for 'other' were averaged across other modes.

Table C.2 Car-sharing substitution in the absence of the City CarShare service in San Francisco. Here, 'car' mode aggregates originally reported on-road modes, and the reported public transportation distance has been split equally into 'train' and 'bus' modes as they have not been distinguished. Source: (Cervero et al., 2007)

| Mode of transport | Kilometres (in %) |
|---|---|
| Car | 27.3 |
| Train | 14.3 |
| Bus, tram, rapid transit | 14.3 |
| Bicycle | 3.9 |
| Walking | 6.9 |
| Other | 3.2 |
| Not travelled | 30.1 |

Calgary

Given driving distances for the 'before' period, distances for the rest of the 'before' modes are estimated proportionally to the existing official figures on a complete modal breakdown in Calgary (Behan and Lea, 2014), see Table C.3. Next, the resulting before-and-after total distance gap was redistributed between the alternative modes in the 'during' period according to the same profile. This allowed accounting for a change in modal distances caused by CS participation ('before' and 'during' distances in Table 5). Per-PKT emissions for 'other' mode were averaged across other modes. Life-cycle electricity-related emissions factor for the province of Alberta in Canada was used (see Appendix D).

Table C.3 Modal breakdown for Calgary. Numbers are rounded based on the original source. The reported public transportation ration has been split equally into 'train' and 'bus' modes as they have not been distinguished. 'Walk and bicycle' mode has been equally split into two separate modes as well. Source: (Behan and Lea, 2014)

| Mode of transport | Kilometres (in %) |
|---|---|
| Car | 76 |
| Train | 8 |
| Bus | 8 |
| Bicycle | 4 |
| Walking | 4 |

Appendix D: Lifetime mileage analysis

A correlation between end of life vehicle (ELV) state of a vehicle and the total mileage or the life time parameters would suggest that these parameters could predict the ELV state, hence allowing to estimate the LTM of shared vehicles given the intensity of their use.

For this purpose, the UK's periodic technical inspection (MOT) vehicle database has been analysed. This database allows to track the same vehicle using their unique vehicle IDs from 2013 to 2015. As soon as a vehicle appeared at the test for 2013, failed it for that year, and never appeared back for the test within next two years in the database, the vehicle was considered to be discarded. It was shown in a similar study that such an approach mitigates distortions because of the crashed or exported vehicles (Dun et al., 2015).

In total, 156,838 ELVs were extracted from the 2013 dataset with an average age of 14.7 years and average mileage of 173,000 km. These data were balanced with non-ELVs from the same year to prepare for a logistic regression analysis. In particular, two models with one independent parameter each were assessed - the age and the mileage of the vehicle, to predict the vehicle's end-of-life status.

As a result (Table D.1), even though both parameters had a statistically significant positive relationship with the ELV status (positive regressions coefficients and P-values lower than 0.05), none of them explained the variance of the dependent variable well enough (very low pseudo R-squared values).

Table D.1 Logistic regression analysis results. ELV binary status – as a dependent variable. The age or the total mileage are the independent variables for two models.

| Model | Coef. | Std. err. | z | P | Pseudo R2 |
|---|---|---|---|---|---|
| age | 0.0329 | 0.0029 | 11.5040 | 0.0000 | 0.03 |
| mileage (1000 km) | 0.0022 | 0.0002 | 9.5918 | 0.0000 | 0.02 |

Such results do not allow to prove any hypothesis about explanatory power of the age and mileage for the end-of-life of the automobile.

Appendix E: Car2Go fleet lifetime mileage

Table E.1 depicts the fleet sizes (Car2Go: Pioneer And Market Leader In Free-Floating Carsharing, 2017) and the total annual Car2Go mileage for each city (Martin and Shaheen, 2016).

Table E.1 Estimating car-sharing fleet's lifetime mileage-based Car2Go data for four cities. VKT = Vehicle kilometres travelled. Distance units have been made converted to km.

|  | Calgary | Seattle | Vancouver | Washington D.C | Average |
|---|---|---|---|---|---|
| Total VKT (annual, km) | 8 400 000 | 9 900 000 | 9 108 000 | 14 700 000 |  |
| Fleet size | 630 | 750 | 1275 | 600 |  |
| Mileage (annual, km) | 13 300 | 13 100 | 11 500 | 9 700 | 11 900 |

A recent Car2Go press release reported 90 million kilometres driven by 14,000 Car2Go vehicles in 6 months (Car2Go: Press Release, 2018). This translates to the 12900 km in annual mileage for car-sharing vehicles. Thus, combining this with the average result from the Table E.1, it could be assumed for this scenario that relatively lower annual distances (12200 km) are driven by the shared automobiles during the same lifetime (15 years).

Appendix F: Existing misalignments

Interestingly, the Dutch study assumed a constant 15 years lifetime for shared and private vehicles and the LTM of 250,000 km (Nijland and van Meerkerk, 2017) for the manufacturing emissions calculations; however, it quantifies vehicle's manufacturing emissions indirectly based on a study with a very different automobile LTM assumed, namely 150,000 km (Nijland and van Meerkerk, 2017). If calculated given the corresponding LTM, the reduction in manufacturing emissions in their study would drop from the proposed 125-281 to 142-186 kg $CO_2$-eq per year. This exemplifies how sensitive the total results are to the LTM assumptions, and suggests another explanation for the lower emission savings from CS in this study.

Nijland and van Meerkerk (2017) mistakenly multiply the average amount of the reduced vehicles per average CS member (-0.4) caused by participation by the shared car 'ownership' (1/15 portion of the shared vehicle from the platform) instead of adding these numbers. Technically, the overall vehicle ownership change after selling the vehicle and 'owning' a portion of the CS fleet has to be: -0.4 + 1/15 = -0.3(3). Moreover, their study quantifies vehicle's manufacturing emissions indirectly through their proposed LTM (250 000 km) based on another study (Gbeghaje-Das, E., 2013) with a very different LTM assumed (150 000 km). Accounting for only these two last misconceptions, the reductions in manufacturing emissions drop from the proposed 125-281 to 142-186 kg CO2e per year which is an enormous difference relative to their final results of 236−392 kg CO2e per year reduction for the average car-sharing member.

Appendix G: Private vehicle's lifetime mileage estimation

Table G.1 Lifetime and lifetime mileage for an average private automobile in the USA. Reported by different studies and their average. Distance units have been made converted to km.

| Source | Age (years) | Annual mileage (km) | $LTM_{car}$ (km) |
|---|---|---|---|
| (Mitropoulos and Prevedouros, 2014) | 10.6 | 18 100 | 192000 |
| (Martin et al., 2010) | 17.3 | 12 300 | 213 000 |
| (Chester, 2008) | 16.9 | 17 600 | 298 000 |
| Average | 15.0 | 16 000 | 234 000 |